\documentclass{INTERSPEECH2023}

\usepackage{booktabs}
\usepackage{cite}
\usepackage{subcaption}

\interspeechcameraready

\title{Speaker Embeddings as Individuality Proxy for Voice Stress Detection}
\name{Zihan Wu$^{1, 2}$\thanks{$^*$ZW performed this work as an intern at Logitech.}, Neil Scheidwasser-Clow$^3$, Karl El Hajal$^{1}$, Milos Cernak$^2$}
\address{
  $^1$École Polytechnique Fédérale de Lausanne (EPFL), Lausanne, Switzerland\\
  $^2$Logitech Europe S.A., Lausanne, Switzerland\\
  $^3$University of Copenhagen, Copenhagen, Denmark}
\email{zihan.wu@epfl.ch, milos.cernak@ieee.org}

\begin{document}

\maketitle
 
\begin{abstract}
Since the mental states of the speaker modulate speech, stress introduced by cognitive or physical loads could be detected in the voice. The existing voice stress detection benchmark has shown that the audio embeddings extracted from the Hybrid BYOL-S self-supervised model perform well. However, the benchmark only evaluates performance separately on each dataset, but does not evaluate performance across the different types of stress and different languages. Moreover, previous studies found strong individual differences in stress susceptibility. This paper presents the design and development of voice stress detection, trained on more than  100 speakers from 9 language groups and five different types of stress. We address individual variabilities in voice stress analysis by adding speaker embeddings to the hybrid BYOL-S features. The proposed method significantly improves voice stress detection performance with an input audio length of only 3–5 seconds. 
\end{abstract}


\section{Introduction}

Recent developments in neural network-based models of speech have spearheaded progress in a variety of discriminative paralinguistic tasks, e.g., speaker count identification~\cite{niizumi2022}, verification~\cite{resemblyzer}, pitch estimation~\cite{kim2018} and emotion recognition~\cite{scheidwasser2022}. Beyond these tasks, robust representations of speech have garnered interest for their potential applications in healthcare, including speech enhancement of hearing aids~\cite{graetzer2021, lesica2021}, speech pathology, and stress assessment~\cite{elbanna22_interspeech, janbakhshi2022, violeta22}. 

Regarding stress assessment, a substantial body of literature has been dedicated to voice stress analysis. The categorized stressor types notably include cognitive load (i.e., task-induced mental effort)~\cite{schuller2013interspeech, jing2014, schuller2014interspeech, van2014, puyvelde2018, elbanna22_interspeech}, physical load (task-induced physical effort)~\cite{jing2014, schuller2014interspeech, elbanna22_interspeech}, emotional load and alcohol/sleep-deprivation/hypoxia~\cite{puyvelde2018}. 
Nevertheless, inter-individual variability remains a common obstacle to producing robust and personalized evaluations~\cite{giddens2013}. There has been remarkably little scientific research to explore the voice as a potential stress indicator, caused by the high diversity of the different types of stress and previously mentioned significant individual differences in stress susceptibility~\cite{scherer2018}.

In addition, several of the proposed models process medium to large utterances (10-30 s~\cite{gallardo2019, schuller2014interspeech, elbanna22_interspeech}) while real-time evaluation is a desired feature for medical or commercial usage.
Furthermore, in a recent benchmark evaluation~\cite{elbanna22_interspeech}, each of the five datasets is evaluated by a separate classifier. However, different datasets contain different languages and different types of cognitive or physical loads. As a result, the classifiers in the benchmark evaluations could be biased toward a specific dataset and thus not suitable for practical usage. It has been shown that cross-corpus testing performances deteriorate in many speech-related tasks such as speech emotion recognition~\cite{zhang21_crosscorpus, schuller10_ieee}. 

This paper also addresses individual variabilities in voice stress analysis. Recent work showed a significant relationship exists between a listener's perceived task load and their personality and frustration intolerance, which were measured for each individual through questionnaires~\cite{personalized_task_load_prediction}. Indeed, the effect of a stimulus on an individual's perception can be highly dependent on individuality factors such as their personality traits and current emotional states~\cite{bias_personality_analysis}. For instance, an easily frustrated listener will perceive tasks as more demanding than one with a higher tolerance for frustration. Consequently, extracting information about a speaker's individuality can enable a more accurate prediction of their perceived cognitive load. We, therefore, hypothesize that extracting speaker identity embeddings, which act as a proxy for individuality factors, can improve the prediction of a speaker's perceived cognitive load. If our hypothesis were correct, it would pave the way for more general personalized paralinguistic speech processing.

The goal of this paper is twofold:
\begin{enumerate}
    \item Building a system that performs well with combined voice stress analysis datasets (using cross-dataset settings as confirmation of better generalization) and works well with short, 3–5 seconds long, input audio clips (some recordings used in~\cite{elbanna22_interspeech} are more than five minutes long!).
    \item Evaluating the usage of speaker embeddings for addressing individual variabilities in voice stress analysis.
\end{enumerate}

More specifically, in this paper, we re-evaluate the performance of Hybrid BYOL-S/CvT~\cite{byols-elbanna22a} 
as a general-purpose speech representation for \textit{personalized} cognitive and physical load detection using the existing cognitive and physical load datasets. While~\cite{elbanna22_interspeech} showed reasonable performance after training on each dataset individually, cross-training across combined datasets markedly deteriorated prediction performance. Further performance decrease is found by limiting the input length of recordings. To address the individual variabilities, we concatenate the pre-trained audio representation~\cite{byols-elbanna22a} with well-established speaker embeddings such as Resemblyzer~\cite{resemblyzer} and ECAPA~\cite{ecapa} and evaluated them on the 3-5-s long cross-dataset voices stress data. We conclude that personalized audio representation significantly improves voice stress detection. 

The paper is structured as follows: Section~\ref{sec:methods} presents the proposed method of personalized model for voice stress analysis. Section~\ref{sec:results} describes the experimental setup and discusses the obtained results. Finally, Section~\ref{sec:conclusion} concludes the paper and suggests future work.

\section{Methods}
\label{sec:methods}

\subsection{Related work}
\label{sec:byols}

The hybrid BYOL-S/CvT~\cite{byols-elbanna22a}, a self-supervised model pre-trained to extract features from speech audio input, was found as a state-of-the-art feature extractor for cognitive and physical task load detection~\cite{elbanna22_interspeech}. The model is an extension of the BYOL-A model~\cite{niizumi2021byol}. During pre-training, the model has two branches of neural networks, each receiving a different augmented version of the input log-mel spectrogram. The model aims to minimize the self-supervised loss: mean-squared error loss between the outputs of the two branches. In this way, the neural networks could capture the meaningful representations underlying the two differently augmented inputs. Both networks have an encoder and a projection module that share the same architectures across the networks, while one branch has an additional predictor to avoid collapsed representations. 

In contrast to BYOL-A which is trained on AudioSet~\cite{gemmeke2017audio}, BYOL-S is solely trained on the subset containing speech data. Besides, BYOL-S/CvT changed the convolutional neural network encoder in BYOL-S to a shallow convolutional vision transformer (CvT)~\cite{wu2021}. The hybrid BYOL-S/CvT model adds a supervision loss that computes the difference between the predictor output and the openSMILE ComParE feature set~\cite{eyben2010opensmile}, which comprises 6373 hand-engineered, DSP-based, audio features.

After pre-training, the CvT encoder of the hybrid BYOL-S/CvT model can be used to extract the embedding features from audio data. A log-mel spectrogram is first computed for each audio sample using a 25 ms window size and a 10 ms hop size. Subsequently, the encoder outputs one 2048-dimensional embedding per 160-ms frame. The utterance-level embeddings can be computed by adding a 'mean+max' temporal pooling. The embedding features extracted by the hybrid BYOL-S/CvT model have shown good performance on HEAR benchmark~\cite{turian2022}, which uses the extracted features in various downstream tasks, including speech, music, and environment-related tasks.

\subsection{Utterance-level and  framewise audio embeddings}

Traditional methods for speech feature extraction mostly obtain a single feature vector for a given utterance~\cite{elbanna22_interspeech, eyben2010opensmile}. Such utterance-level embeddings condensed the temporal information stored in sequential data like speech. In contrast, framewise embeddings are obtained by using a model to extract a sequence of embeddings from sequential data. Such embeddings preserve the temporal information stored in the input data. Meanwhile, advancements in deep neural networks have proposed models like Transformers~\cite{ashish2017_attention} that are capable of capturing temporal information in framewise embeddings. Framewise embeddings have been shown to outperform utterance-level embeddings in speech quality assessment tasks~\cite{mosra-byols}. 

In this work, we extracted both types of embeddings and evaluated their performance in voice stress detection tasks. The utterance-level embeddings were extracted by using “mean+max” pooling on the Hybrid BYOL-S/CvT model output (Section~\ref{sec:byols}), and the framewise embeddings were extracted by removing that 'mean+max' pooling layer. 

\subsection{Combined audio and speaker embeddings}

\begin{figure}[t]
  \centering
  \includegraphics[width=\linewidth]{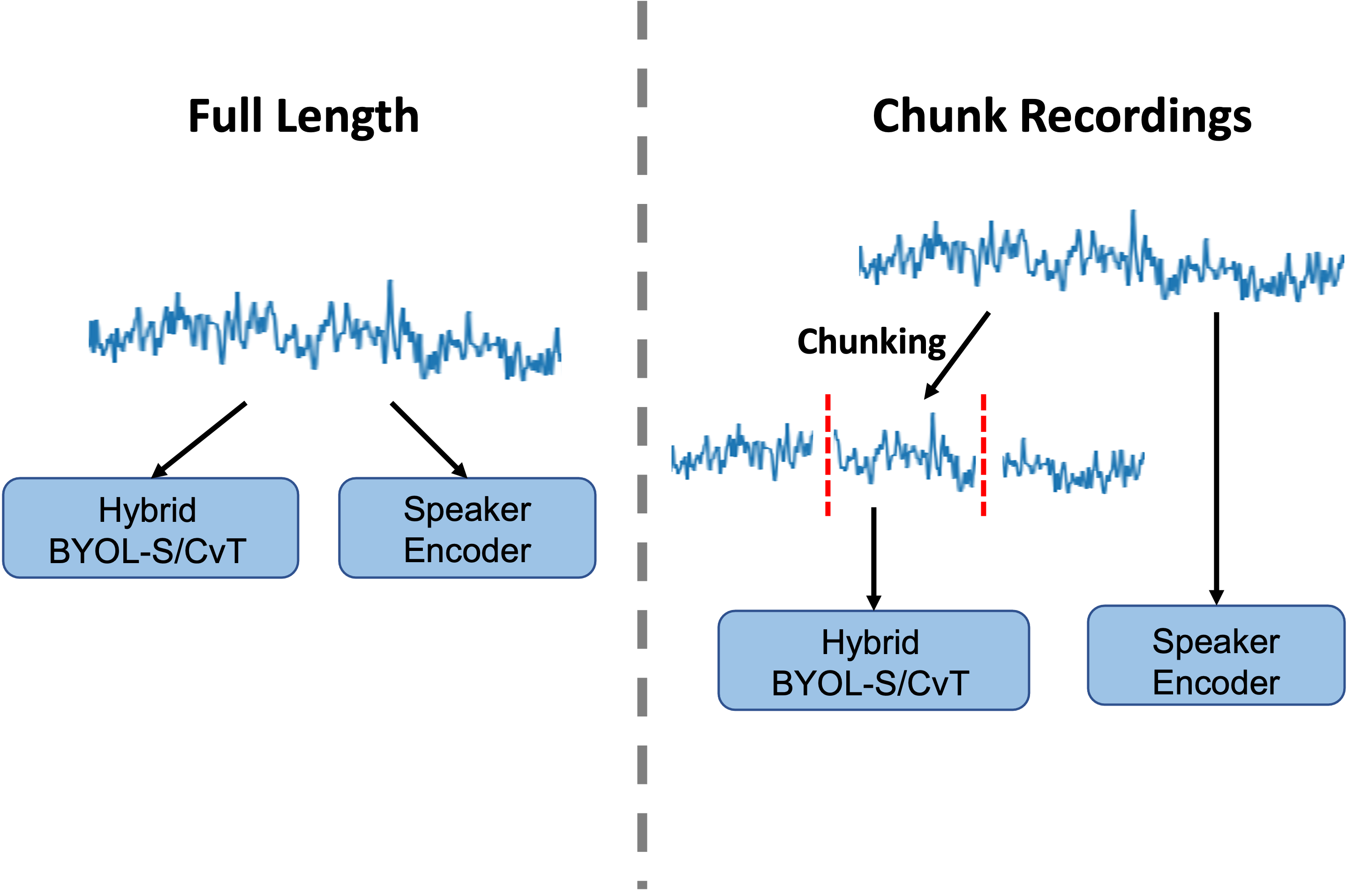}
  \caption{Left: Full-length audio embedding extraction.  Right: Chunked audio embedding extraction.}
  \label{fig:chunk}
\end{figure}

\begin{figure}[t]
  \centering
  \includegraphics[width=\linewidth]{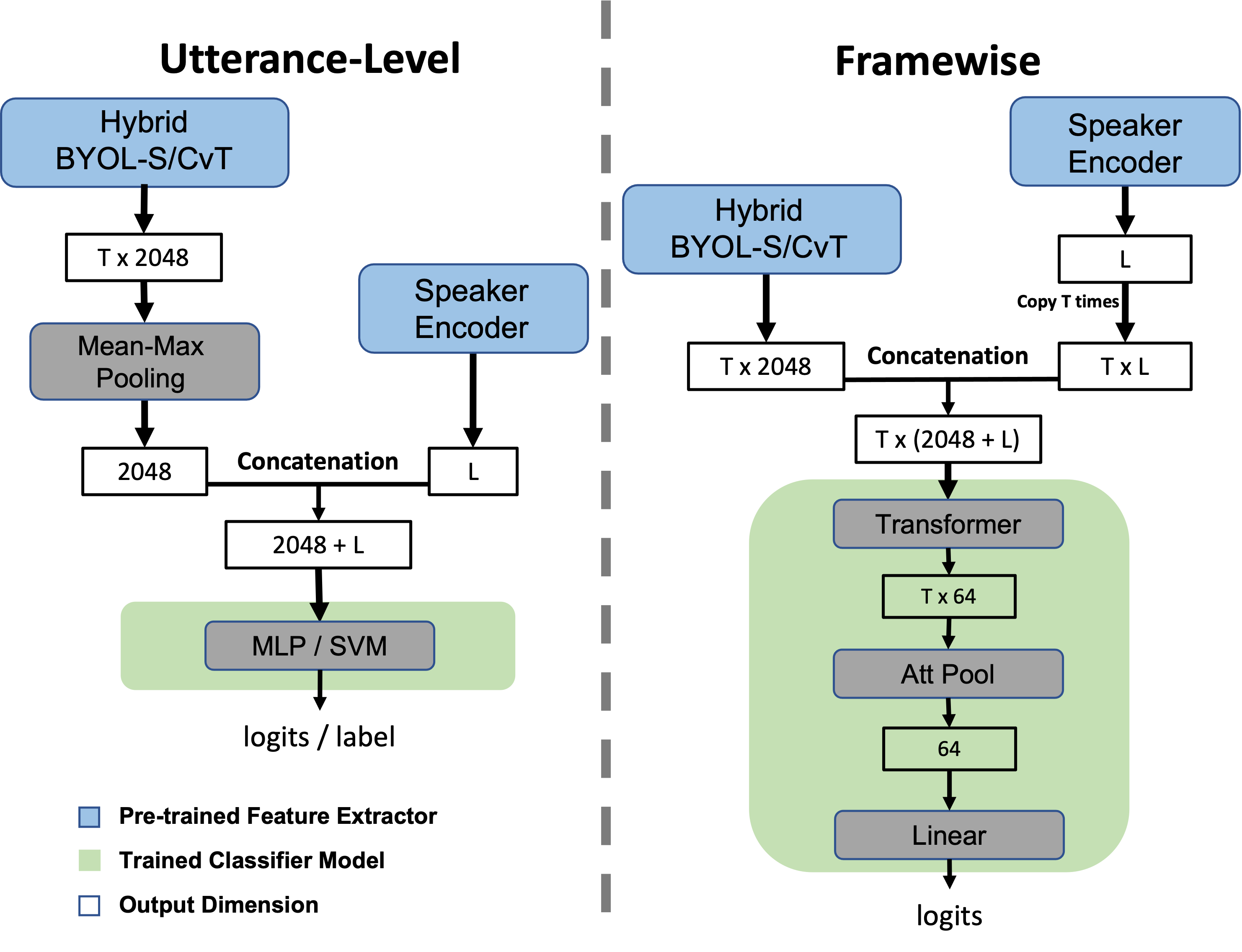}
  \caption{Concatenation of speaker embeddings and model embeddings. Left: Utterance-level embedding concatenation; Right: Framewise embedding concatenation.}
  \label{fig:concat}
\end{figure}

We propose to use speaker embeddings as a proxy for the individuality factors like personality and frustration intolerance, which are important for cognitive load perception~\cite{personalized_task_load_prediction}.
In this work, we considered two speaker encoders to extract speaker embeddings: Resemblyzer~\cite{resemblyzer} and ECAPA~\cite{ecapa}. Both encoders generate a speaker embedding vector for each speech utterance. Resemblyzer embeddings have 256 features, and ECAPA embeddings have 196 features.

Fig.~\ref{fig:chunk} shows full-length and chunked audio speaker embedding computation. When using full-length data, the speaker embeddings and the audio embeddings were concatenated. When using chunked (3-5 seconds long) data, the audio embeddings were computed using only the chunked audio clips. Each chunk has the same target label as its original recording. The audio embeddings of the chunked audio clip and the speaker embeddings of the full-length utterance were then concatenated and used for training downstream classifiers. We did not compute separate speaker embeddings using each chunked clip because the whole utterance better represents the speaker's identity. Clips chunked from the same utterance share the same speaker embeddings.

\begin{table*}[thbp]
  \caption{Unweighted Average Recall (UAR) on the test set. Score: mean $\pm$ 95\% bootstrapped confidence interval (CI)}
  \label{tab:speaker_embed}
  \centering
  \begin{tabular}{ l c cc cc }
    \toprule
     & \multicolumn{1}{c}{\textbf{full length}} & \multicolumn{1}{c}{\textbf{5-s chunk}} & \multicolumn{1}{c}{\textbf{5-s chunk}} & \multicolumn{1}{c}{\textbf{3-s chunk}} & \multicolumn{1}{c}{\textbf{3-s chunk}}\\
     & \multicolumn{1}{c}{\textbf{utterance-level}} & \multicolumn{1}{c}{\textbf{utterance-level}} & \multicolumn{1}{c}{\textbf{framewise}} & \multicolumn{1}{c}{\textbf{utterance-level}} & \multicolumn{1}{c}{\textbf{framewise}}\\
    \midrule
    \textbf{\textit{Only audio embeddings}}: & & & & & \\
    Hybrid BYOL-S/CvT  & 74.99 $\pm$ 2.88 & 70.43 $\pm$ 1.27 & 70.66 $\pm$ 1.22 & 67.00 $\pm$ 1.10 & 67.65 $\pm$ 1.06\\
    Hybrid BYOL-S/CvT ($\alpha$ = 2)  & \textbf{77.07} $\pm$ 2.82 & 70.20 $\pm$ 1.27 & 68.54 $\pm$ 1.30 & 70.18 $\pm$ 1.04 & 70.73 $\pm$ 1.06  \\
    \midrule 
    \textbf{\textit{Only speaker embeddings}}: & & & & & \\
    ECAPA & 63.16 $\pm$ 3.24 & 75.73 $\pm$ 1.26 & -$^*$ & 75.49 $\pm$ 1.04 & -\\
    Resemblyzer & 66.45 $\pm$ 3.31 & 70.57 $\pm$ 1.32 & -\ \  & 69.94 $\pm$ 1.01 & -\\
    ECAPA + Resemblyzer$^1$ & 62.41 $\pm$ 3.12 & 74.23 $\pm$ 1.28 & -\ \  & 74.55 $\pm$ 0.98 & -\\
    \midrule
    \textbf{\textit{Concatenate  ECAPA}}: & & & & & \\
    Hybrid BYOL-S/CvT & 72.93 $\pm$ 2.95 & \textbf{80.21} $\pm$ 1.16 & \textbf{79.32} $\pm$ 1.10 & \textbf{80.62} $\pm$ 0.96 & 79.59 $\pm$ 0.86\\
    Hybrid BYOL-S/CvT ($\alpha$ = 2)  & 76.20 $\pm$ 2.74 & 79.24 $\pm$ 1.10  & 78.25 $\pm$ 1.12 & 79.38 $\pm$ 0.94  & \textbf{80.89} $\pm$ 0.90   \\
    \textbf{\textit{Concatenate Resemblyzer}}: & & & & & \\
    Hybrid BYOL-S/CvT  & 76.55 $\pm$ 2.89 & 72.08 $\pm$ 1.26 & 68.52 $\pm$ 1.34 & 71.67 $\pm$ 0.98 & 70.32 $\pm$ 0.96\\
    Hybrid BYOL-S/CvT ($\alpha$ = 2)  & 75.87 $\pm$ 2.86 & 74.50 $\pm$ 1.27 & 71.26 $\pm$ 1.32 & 74.68 $\pm$ 1.02 & 72.83 $\pm$ 1.02  \\
    \bottomrule
    \multicolumn{6}{l}{\small *The speaker embeddings are utterance-level embeddings, so there is no framewise performance for speaker embedding only.}\\
    \multicolumn{6}{l}{\small $^1$The two speaker embeddings are concatenated}
  \end{tabular}
\end{table*}

Fig.~\ref{fig:concat}, left, shows the concatenation of the speaker and audio utterance-level embeddings. After concatenation, we used either Support Vector Machine (SVM) or Multi-Layer Perceptrons (MLP) for binary classification. Fig.~\ref{fig:concat}, right, shows the concatenation of the speaker and audio framewise embeddings. The dimension of audio framewise embeddings is $T \times 2048$, where $T$ is the total audio length divided by 160ms. We concatenated the same utterance-level speaker embedding at each timestamp. As a result, the framewise embeddings had the shape $T \times (2048 + L)$ where $L$ is the length of speaker embedding. After concatenation, we used a lightweight Transformer followed by an attention-pooling layer and a linear projection layer for binary classification.

\section{Experimental setup and results}
\label{sec:results}
\subsection{Data}

We used the existing cognitive and physical load dataset corpus as presented in~\cite{elbanna22_interspeech}. The corpus consists of five datasets, with a total number of 111 speakers, nine languages, and an overall duration of 12 hours. For each data sample, the task is to predict whether the recorded speech was produced while the speaker was simultaneously performing a cognitive or physical load-inducing task or not.


In each one of the five datasets, the audio samples were split based on speaker identities into 60\% for training, 15\% for validation, and 25 \% for testing. As a result, the testing data only consisted of speakers never seen by the model during training. 

\subsection{Baseline systems}

To be consistent with the benchmark, we first used the hybrid BYOL-S/CvT model to extract the audio embeddings. In our experiments, we picked two best-performing models from the benchmark paper: the hybrid BYOL-S/CvT model and a variation of the hybrid BYOL-S/CvT model in which the weight of supervision loss to self-supervised loss ($\alpha = \alpha_{sup} : \alpha_{s1} = 2$) is 2:1.

\begin{table}[th]
  \caption{Test UAR after combining datasets. Score: mean $\pm$ 95\% bootstrapped CI}
  \label{tab:combine}
  \centering
  \begin{tabular}{ l cc }
    \toprule
     & \multicolumn{1}{c}{\textbf{Average}$^*$}& \multicolumn{1}{c}{\textbf{Combined}} \\
    \midrule
    Hybrid BYOL-S/CvT & 80.36 & 73.68 $\pm$ 3.10$^1$   \\
    Hybrid BYOL-S/CvT ($\alpha=2$)  & 82.43 & 73.72 $\pm$ 2.93 \ \ \\
    \bottomrule
    \multicolumn{3}{l}{\small *CI not provided because it is an average performance on}\\
    \multicolumn{3}{l}{\small \ \ different datasets}
  \end{tabular}
\end{table}

The utterance-level embeddings were extracted by the hybrid BYOL-S/CvT models with the 'mean+max' pooling layer to compress the temporal dimension of the model output. Then, we used an SVM as the downstream classifier and trained an SVM on each dataset separately. During training, we searched the optimal penalty weight from $10^{-5}$ to $10^5$, and we used the 5-fold cross-validated recall score to determine the optimal weight. Table~\ref{tab:combine} (“average” column) shows the reproduced results of~\cite{elbanna22_interspeech} as an average of the unweighted average recall (UAR) of the five datasets. 


We also see that the UAR performance significantly dropped (“combined” column) after we combined the five datasets. We used the same SVM model and hyperparameter search methods. 

\subsection{Neural network downstream classifier}

To improve performance, we changed the SVM classifier to a shallow MLP, with one hidden layer of 512 neurons. During the training, the classification accuracy of the validation dataset is used for early stopping, with 30 epochs of patience. We also used the validation performance to search the optimal learning rate from $ 3.2\times10^{-3}, 10^{-3}, 3.2\times10^{-4}, 10^{-4}, 3.2\times10^{-5}$.

As shown in Table~\ref{tab:speaker_embed} (column “full length” and rows “only audio embedding”), MLP-based classification improved performance on the combined dataset.

\subsection{Decreasing input audio length}

To train a model capable of making predictions based on shorter audio inputs, we first chunked all audio data to five seconds. Then, we used the hybrid BYOL-S/CvT models to extract the utterance-level embeddings on these shorter inputs and searched for the best downstream classifier among the SVM and MLP. The training process for SVM and MLP was the same as before chunking. The performance decreased significantly after chunking the data (Table~\ref{tab:speaker_embed}, 5-s chunk utterance-level column and only audio embedding rows).

\subsection{Utterance-level speaker embeddings}

\subsubsection{Combined with audio embeddings from 5-s chunked data} 

Our next step was to address individual variabilities in voice stress analysis by using speaker embeddings. We chose two speaker embeddings encoders: ECAPA~\cite{ecapa} (implemented by speechbrain~\cite{speechbrain}) and Resemblyzer~\cite{resemblyzer} (implemented by resemble-ai\footnote{\href{https://github.com/resemble-ai/Resemblyzer}{https://github.com/resemble-ai/Resemblyzer}}).
After concatenation of ECAPA speaker embeddings and utterance-level audio embeddings, performance increases significantly, with ECAPA and hybrid BYOL-S/CvT embeddings reaching a test UAR of 80.21\% (Table~\ref{tab:speaker_embed}, 5-s chunk utterance-level column and concatenate ECAPA rows). Interestingly, we found Resemblyzer embeddings give much less improvement on the performance than ECAPA does (Table~\ref{tab:speaker_embed}, 5-s chunk utterance-level column and concatenate Resemblyzer rows). 

Besides, we tried to add speaker embeddings to the audio embeddings of the original full-length dataset. The speaker embeddings, however, do not always help improve the model performance (Table~\ref{tab:speaker_embed}, full-length utterance-level column). 

\subsubsection{Speaker embeddings only} 

To investigate the differences between the two speaker encoders, we evaluated the performance using only speaker embeddings. For each chunked data, the speaker embeddings were still computed from the whole utterance before chunking. We found ECAPA has a better performance than Resemblyzer (Table~\ref{tab:speaker_embed}, “only speaker embeddings” rows). This matches the performance gap between concatenating ECAPA embeddings and concatenating Resemblyzer embeddings. Interestingly, for chunked data, we found that ECAPA embeddings themselves even outperformed the embeddings from the two hybrid BYOL-S models. We speculated that the speaker embedding models themselves contained not only information on speaker identities, but also some paralinguistic features that help the downstream model to detect the cognitive or physical loads. The additional paralinguistic information stored in the ECAPA embeddings helped to improve the classification performance. We also observed significant improvement when using speaker embeddings only on 5-s chunks instead of full-length utterance-level data. We believe it was due to having more balanced chunked data -- in the full-length version, for example, several minutes long audio had a single label, whereas with chunked data the label distribution became more balanced.

Furthermore, we examined whether the improvement brought by ECAPA concatenation is simply due to the additional paralinguistics information in the ECAPA embeddings. Thus, we evaluated the concatenation of ECAPA and Resemblyzer embeddings. The obtained performance does not exceed the better performance of ECAPA-only on chunked data (Table~\ref{tab:speaker_embed}, only speaker embedding rows). In this case, adding additional information does not lead to performance incease. Consequently, we argue that the speaker identity and paralinguistics information encoded by ECAPA really complements the audio features extracted by the hybrid BYOL-S/CvT in voice stress detection tasks.

\subsubsection{Combined with audio embeddings from 3-s chunked data}
Finally, we examined the sensitivity of the different models to shorter chunk sizes (3 instead of 5 s). Regarding utterance-level embeddings, performance of hybrid BYOL-S/CvT $(\alpha = 2)$ showed little variation, while the hybrid BYOL-S/CvT showed worse performance on shorter audio chunks. After concatenating the 3-s chunked utterance-level embeddings with the speaker embeddings, both models showed a performance similar to 5-s chunking length. Concatenation of speaker embeddings helped the models to be more robust to different chunking lengths.

\subsection{Framewise audio embeddings}

We have reported so far only the results for the utterance-level embeddings, whereas framewise embeddings have also been used to improve the performance of many speech-related tasks like speech quality assessment~\cite{mosra-byols}. With chunked data, we were able to get framewise embeddings with shorter and less varied temporal lengths. Therefore, we also extracted framewise embeddings and trained a two-layer, 64-dimensional, Transformer-based classifier on the embeddings to detect the voice stress.

We also used validation accuracy with 30 epochs of patience for early stopping. The optimal learning rate was again searched from $ 3.2\times10^{-3}, 10^{-3}, 3.2\times10^{-4}, 10^{-4}, 3.2\times10^{-5}$.

On 5-s chunked data, using framewise embeddings did not improve the recall performance, regardless of speaker embedding concatenation (Table~\ref{tab:speaker_embed}, 5-s chunk utterance-level columns and concatenate ECAPA or Resemblyzer rows). When we chunked the data to 3 s, the performance improves slightly for the hybrid BYOL-S/CvT ($\alpha = 2$) model, but not for the hybrid BYOL-S/CvT model. The effect of using framewise embeddings is thus model-dependent. We obtained the best performance by concatenating the hybrid BYOL-S/CvT ($\alpha = 2$) model embeddings and ECAPA embeddings. However, the slight improvement was not significant when compared with the best utterance-level embeddings. Overall, we found framewise embedding as being not helpful in voice stress detection tasks.



\section{Conclusion}
\label{sec:conclusion}
In this work, we have constructed a pipeline to use the embeddings of a pre-trained audio feature extraction model in a cross-dataset, lower latency voice stress detection system. We re-evaluated the state-of-the-art pre-trained model from the previous benchmark paper~\cite{elbanna22_interspeech} and found a worse performance in the new pipeline. We improved the cross-dataset performance by replacing the downstream SVM classifier with a shallow MLP or Transformer. We also confirmed that individual variabilities in voice stress analysis can be achieved by concatenating ECAPA speaker embeddings, which significantly improved the performance after we chunked the data into shorter segments.

In our future work, we aim to include other types of stressors, such as emotional load, to train a more general voice stress analysis model. Besides, we wish to apply the speaker embeddings in other speech-related downstream tasks, including emotion recognition and quality assessments, to examine the role of individuality factors in these tasks. Finally, it is also worthy of investigating the relationship between speaker identity and paralinguistic features to help explain the good performance of speaker embeddings on this voice stress detection task.

\section{Acknowledgements}

We would like to thank Gasser Elbanna and Pierre Beckmann who offered their critical discussions and suggestions on how to well validate the proposed method.



\bibliographystyle{IEEEtran}
\bibliography{References}

\begin{thebibliography}{10}
\providecommand{\url}[1]{#1}
\csname url@samestyle\endcsname
\providecommand{\newblock}{\relax}
\providecommand{\bibinfo}[2]{#2}
\providecommand{\BIBentrySTDinterwordspacing}{\spaceskip=0pt\relax}
\providecommand{\BIBentryALTinterwordstretchfactor}{4}
\providecommand{\BIBentryALTinterwordspacing}{\spaceskip=\fontdimen2\font plus
\BIBentryALTinterwordstretchfactor\fontdimen3\font minus
  \fontdimen4\font\relax}
\providecommand{\BIBforeignlanguage}[2]{{%
\expandafter\ifx\csname l@#1\endcsname\relax
\typeout{** WARNING: IEEEtran.bst: No hyphenation pattern has been}%
\typeout{** loaded for the language `#1'. Using the pattern for}%
\typeout{** the default language instead.}%
\else
\language=\csname l@#1\endcsname
\fi
#2}}
\providecommand{\BIBdecl}{\relax}
\BIBdecl

\bibitem{niizumi2022}
D.~Niizumi, D.~Takeuchi, Y.~Ohishi, N.~Harada, and K.~Kashino, ``{Masked
  Spectrogram Modeling using Masked Autoencoders for Learning General-purpose
  Audio Representation},'' in \emph{HEAR: Holistic Evaluation of Audio
  Representations (NeurIPS 2021 Competition)}, vol. 166, 2022, pp. 1--24.

\bibitem{resemblyzer}
L.~Wan, Q.~Wang, A.~Papir, and I.~L. Moreno, ``Generalized end-to-end loss for
  speaker verification,'' in \emph{ICASSP}, 2018, p. 4879–4883.

\bibitem{kim2018}
J.~W. Kim, J.~Salamon, P.~Li, and J.~P. Bello, ``Crepe: A convolutional
  representation for pitch estimation,'' in \emph{ICASSP}, 2018, pp. 161--165.

\bibitem{scheidwasser2022}
N.~Scheidwasser-Clow, M.~Kegler, P.~Beckmann, and M.~Cernak, ``Serab: A
  multi-lingual benchmark for speech emotion recognition,'' in \emph{ICASSP},
  2022, pp. 7697--7701.

\bibitem{graetzer2021}
S.~Graetzer, J.~Barker, T.~J. Cox, M.~Akeroyd, J.~F. Culling, G.~Naylor,
  E.~Porter, R.~Viveros~Munoz \emph{et~al.}, ``Clarity-2021 challenges: Machine
  learning challenges for advancing hearing aid processing,'' in \emph{Proc.
  Interspeech}, 2021, pp. 686--690.

\bibitem{lesica2021}
N.~A. Lesica, N.~Mehta, J.~G. Manjaly, L.~Deng, B.~S. Wilson, and F.-G. Zeng,
  ``Harnessing the power of artificial intelligence to transform hearing
  healthcare and research,'' \emph{Nat. Mach. Intell.}, vol.~3, no.~10, pp.
  840--849, 2021.

\bibitem{elbanna22_interspeech}
G.~Elbanna, A.~Biryukov, N.~Scheidwasser-Clow, L.~Orlandic, P.~Mainar,
  M.~Kegler, P.~Beckmann, and M.~Cernak, ``{Hybrid Handcrafted and Learnable
  Audio Representation for Analysis of Speech Under Cognitive and Physical
  Load},'' in \emph{Proc. Interspeech}, 2022, pp. 386--390.

\bibitem{janbakhshi2022}
P.~Janbakhshi and I.~Kodrasi, ``Experimental investigation on stft phase
  representations for deep learning-based dysarthric speech detection,'' in
  \emph{ICASSP}, 2022, pp. 6477--6481.

\bibitem{violeta22}
L.~P. Violeta, W.~C. Huang, and T.~Toda, ``{Investigating Self-supervised
  Pretraining Frameworks for Pathological Speech Recognition},'' in \emph{Proc.
  Interspeech}, 2022, pp. 41--45.

\bibitem{schuller2013interspeech}
B.~Schuller, S.~Steidl, A.~Batliner, A.~Vinciarelli, K.~Scherer, F.~Ringeval,
  M.~Chetouani, F.~Weninger, F.~Eyben, E.~Marchi \emph{et~al.}, ``{The
  INTERSPEECH 2013 Computational Paralinguistics Challenge: Social Signals,
  Conflict, Emotion, Autism},'' in \emph{Proc. Interspeech}, 2013, pp.
  148--152.

\bibitem{jing2014}
H.~Jing, T.-Y. Hu, H.-S. Lee, W.-C. Chen, C.-C. Lee, Y.~Tsao, and H.-M. Wang,
  ``{Ensemble of Machine Learning Algorithms for Cognitive and Physical Speaker
  Load Detection},'' in \emph{Proc. Interspeech 2014}, 2014, pp. 447--451.

\bibitem{schuller2014interspeech}
B.~Schuller, S.~Steidl, A.~Batliner, J.~Epps, F.~Eyben, F.~Ringeval, E.~Marchi,
  and Y.~Zhang, ``{The INTERSPEECH 2014 Computational Paralinguistics
  Challenge: Cognitive \& Physical Load},'' in \emph{Proc. Interspeech 2014},
  2014, pp. 427--431.

\bibitem{van2014}
M.~Van~Segbroeck, R.~Travadi, C.~Vaz, J.~Kim, M.~P. Black, A.~Potamianos, and
  S.~S. Narayanan, ``{Classification of Cognitive Load from Speech using an
  i-vector Framework},'' in \emph{Proc. Interspeech 2014}, 2014, pp. 751--755.

\bibitem{puyvelde2018}
M.~Van~Puyvelde, X.~Neyt, F.~McGlone, and N.~Pattyn, ``Voice stress analysis: A
  new framework for voice and effort in human performance,'' \emph{Front.
  Psychol.}, vol.~9, 2018.

\bibitem{giddens2013}
C.~L. Giddens, K.~W. Barron, J.~Byrd-Craven, K.~F. Clark, and A.~S. Winter,
  ``{Vocal Indices of Stress: A Review},'' \emph{J. Voice}, vol.~27, no.~3, p.
  390.e21–390.e29, 2013.

\bibitem{scherer2018}
K.~R. Scherer, ``{{Acoustic Patterning of Emotion Vocalizations}},'' in
  \emph{{The Oxford Handbook of Voice Perception}}.\hskip 1em plus 0.5em minus
  0.4em\relax Oxford University Press, 2018.

\bibitem{gallardo2019}
A.~Gallardo-Antolín and J.~M. Montero, ``{A Saliency-Based Attention LSTM
  Model for Cognitive Load Classification from Speech},'' in \emph{Proc.
  Interspeech}, 2019, pp. 216--220.

\bibitem{zhang21_crosscorpus}
S.~Zhang, R.~Liu, X.~Tao, and X.~Zhao, ``Deep cross-corpus speech emotion
  recognition: Recent advances and perspectives,'' \emph{Front. Neurorobotics},
  p. 162, 2021.

\bibitem{schuller10_ieee}
B.~Schuller, B.~Vlasenko, F.~Eyben, M.~Wöllmer, A.~Stuhlsatz, A.~Wendemuth,
  and G.~Rigoll, ``Cross-corpus acoustic emotion recognition: Variances and
  strategies,'' \emph{IEEE Transactions on Affective Computing}, vol.~1, no.~2,
  pp. 119--131, 2010.

\bibitem{personalized_task_load_prediction}
R.~P. Spang, K.~E. Hajal, S.~Möller, and M.~Cernak, ``{Personalized Task Load
  Prediction in Speech Communication},'' in \emph{ICASSP}, 2023, pp. 1--5.

\bibitem{bias_personality_analysis}
R.~Principi, C.~Palmero, J.~Junior, and S.~Escalera, ``On the effect of
  observed subject biases in apparent personality analysis from audio-visual
  signals,'' \emph{IEEE Trans. Affect. Comput.}, vol.~12, no.~03, pp. 607--621,
  2021.

\bibitem{byols-elbanna22a}
G.~Elbanna, N.~Scheidwasser-Clow, M.~Kegler, P.~Beckmann, K.~El~Hajal, and
  M.~Cernak, ``{{BYOL-S}: Learning Self-supervised Speech Representations by
  Bootstrapping},'' in \emph{HEAR: Holistic Evaluation of Audio Representations
  (NeurIPS 2021 Competition)}, vol. 166, 2022, pp. 25--47.

\bibitem{ecapa}
B.~Desplanques, J.~Thienpondt, and K.~Demuynck, ``{ECAPA-TDNN:} emphasized
  channel attention, propagation and aggregation in {TDNN} based speaker
  verification,'' in \emph{Proc. Interspeech}, 2020, pp. 3830--3834.

\bibitem{niizumi2021byol}
D.~Niizumi, D.~Takeuchi, Y.~Ohishi, N.~Harada, and K.~Kashino, ``{BYOL for
  Audio: Self-supervised learning for general-purpose audio representation},''
  in \emph{IJCNN}, 2021, pp. 1--8.

\bibitem{gemmeke2017audio}
J.~F. Gemmeke, D.~P. Ellis, D.~Freedman, A.~Jansen, W.~Lawrence, R.~C. Moore,
  M.~Plakal, and M.~Ritter, ``Audio set: An ontology and human-labeled dataset
  for audio events,'' in \emph{ICASSP}, 2017, pp. 776--780.

\bibitem{wu2021}
H.~Wu, B.~Xiao, N.~Codella, M.~Liu, X.~Dai, L.~Yuan, and L.~Zhang, ``Cv{T}:
  Introducing convolutions to vision transformers,'' in \emph{Proc. ICCV},
  2021, pp. 22--31.

\bibitem{eyben2010opensmile}
F.~Eyben, M.~W{\"o}llmer, and B.~Schuller, ``{openSMILE - The Munich Versatile
  and Fast Open-Source Audio Feature Extractor},'' in \emph{Proc. ACM
  Multimedia (MM)}, 2010, pp. 1459--1462.

\bibitem{turian2022}
J.~Turian, J.~Shier, H.~R. Khan, B.~Raj, B.~W. Schuller, C.~J. Steinmetz,
  C.~Malloy, G.~Tzanetakis, G.~Velarde, K.~McNally, M.~Henry, N.~Pinto,
  C.~Noufi, C.~Clough, D.~Herremans, E.~Fonseca, J.~Engel, J.~Salamon,
  P.~Esling, P.~Manocha, S.~Watanabe, Z.~Jin, and Y.~Bisk, ``{HEAR: Holistic
  Evaluation of Audio Representations},'' in \emph{NeurIPS Competitions and
  Demonstrations Track}, 2022, pp. 125--145.

\bibitem{ashish2017_attention}
A.~Vaswani, N.~Shazeer, N.~Parmar, J.~Uszkoreit, L.~Jones, A.~N. Gomez,
  L.~Kaiser, and I.~Polosukhin, ``Attention is all you need,'' in
  \emph{NeurIPS}, vol.~30, 2017.

\bibitem{mosra-byols}
K.~El~Hajal, Z.~Wu, N.~Scheidwasser-Clow, G.~Elbanna, and M.~Cernak,
  ``Efficient speech quality assessment using self-supervised framewise
  embeddings,'' in \emph{ICASSP}, 2023, pp. 1--5.

\bibitem{speechbrain}
M.~Ravanelli, T.~Parcollet, P.~Plantinga, A.~Rouhe, S.~Cornell, L.~Lugosch,
  C.~Subakan, N.~Dawalatabad, A.~Heba, J.~Zhong, J.-C. Chou, S.-L. Yeh, S.-W.
  Fu, C.-F. Liao, E.~Rastorgueva, F.~Grondin, W.~Aris, H.~Na, Y.~Gao, R.~D.
  Mori, and Y.~Bengio, ``{SpeechBrain}: A general-purpose speech toolkit,''
  \emph{arXiv preprint arXiv:2106.04624}, 2021.

\end{thebibliography}

\end{document}